\documentclass[aps,prb,twocolumn,superscriptaddress,showpacs]{revtex4}
\usepackage{graphicx}
\usepackage{longtable}
\usepackage{array}
\usepackage{braket}
\usepackage{hyperref}
\usepackage{subfiles}
\usepackage{physics}
\newcolumntype{L}{>{\centering\arraybackslash}m{2cm}}
\newcolumntype{R}{>{\centering\arraybackslash}m{1.5cm}}
\newcolumntype{K}{>{\centering\arraybackslash}m{1.3cm}}

\def \YBCL{YbCl$_3$}
\def \LUCL{LuCl$_3$}

\begin{document}


\title{Crystal field splitting, local anisotropy, and low energy excitations in the quantum magnet \YBCL \footnote{This manuscript has been authored by UT-Battelle, LLC under Contract No. DE-AC05-00OR22725 with the U.S. Department of Energy.  The United States Government retains and the publisher, by accepting the article for publication, acknowledges that the United States Government retains a non-exclusive, paid-up, irrevocable, world-wide license to publish or reproduce the published form of this manuscript, or allow others to do so, for United States Government purposes.  The Department of Energy will provide public access to these results of federally sponsored research in accordance with the DOE Public Access Plan (http://energy.gov/downloads/doe-public-access-plan).}}

\author{G. Sala}
\affiliation{Neutron Scattering Division, Oak Ridge National Laboratory, Oak Ridge, Tennessee 37831, USA}

\author{M. B. Stone}
\affiliation{Neutron Scattering Division, Oak Ridge National Laboratory, Oak Ridge, Tennessee 37831, USA}

\author{Binod K. Rai}
\affiliation{Materials Science \& Technology Division, Oak Ridge National Laboratory, Oak Ridge, TN 37831, USA}

\author{A. F. May}
\affiliation{Materials Science \& Technology Division, Oak Ridge National Laboratory, Oak Ridge, TN 37831, USA}

\author{D. S. Parker}
\affiliation{Materials Science \& Technology Division, Oak Ridge National Laboratory, Oak Ridge, TN 37831, USA}

\author{G\'abor B. Hal\'asz}
\affiliation{Materials Science \& Technology Division, Oak Ridge National Laboratory, Oak Ridge, TN 37831, USA}

\author{Y. Q. Cheng}
\affiliation{Neutron Scattering Division, Oak Ridge National Laboratory, Oak Ridge, Tennessee 37831, USA}

\author{G. Ehlers}
\affiliation{Neutron Technologies Division, Oak Ridge National Laboratory, Oak Ridge, TN 37831, USA}

\author{V. O. Garlea}
\affiliation{Neutron Scattering Division, Oak Ridge National Laboratory, Oak Ridge, Tennessee 37831, USA}

\author{Q. Zhang}
\affiliation{Neutron Scattering Division, Oak Ridge National Laboratory, Oak Ridge, Tennessee 37831, USA}

\author{M. D. Lumsden}
\affiliation{Neutron Scattering Division, Oak Ridge National Laboratory, Oak Ridge, Tennessee 37831, USA}

\author{A. D. Christianson}
\email{christiansad@ornl.gov}
\affiliation{Materials Science \& Technology Division, Oak Ridge National Laboratory, Oak Ridge, TN 37831, USA}

\date{\today}

\begin{abstract}
We study the correlated quantum magnet, \YBCL{}, with neutron scattering, magnetic susceptibility, and  heat capacity measurements. The crystal field Hamiltonian is determined through simultaneous refinements of the inelastic neutron scattering and magnetization data. The ground state doublet is well isolated from the other crystal field levels and results in an effective spin-1/2 system with local easy plane anisotropy at low temperature.  Cold neutron spectroscopy shows low energy excitations that are consistent with nearest neighbor antiferromagnetic correlations of reduced dimensionality.
\end{abstract}




\maketitle

The Quantum Spin Liquid (QSL) is a state of matter hosting exotic fractionalized excitations and long range entanglement between spins with potential applications for quantum computing~\cite{Savary_2016, zhou_spinliqrev_2017, Takagi2019, knolle_2019}. Since QSL physics relies on quantum fluctuations that are enhanced by low spin and low dimensionality, spin-1/2 systems on two-dimensional lattices provide a natural experimental platform for realizing a QSL phase. It has also been shown that an \textit{effective} spin-1/2 system can be generated even in compounds with high-angular-momentum ions like Yb$^{3+}$ and Er$^{3+}$, where the combination of crystal-field effects and strong spin-orbit coupling lead to highly anisotropic interactions between effective spin-1/2 degrees of freedom~\cite{Ross2014}. 

Magnetic frustration plays a central role in stabilizing QSL phases~\cite{Balents_2010}. While QSLs were traditionally associated with geometrically frustrated systems (e.g., triangular and kagome lattices), it has recently become well appreciated that \textit{exchange frustration} due to highly anisotropic spin interactions can also stabilize QSL phases, even on bipartite lattices~\cite{Witczak_2014, Rau_2016_review}. Most famously, bond-dependent spin interactions on the honeycomb lattice give rise to the Kitaev model, an exactly solvable model with a gapless QSL ground state~\cite{KITAEV2006}. A number of honeycomb materials, primarily containing 4d or 5d transition metals such as Ir or Ru have been put forth as realizations of the Kitaev model~\cite{Jackeli_2009, Chaloupka_2010}.  Prominent examples include (Na,Li)$_2$IrO$_3$~\cite{singh2010antiferromagnetic, liu2011longrange, singh2012relevance, choi2012spin, ye2012direct, comin2012novel, chun2015direct, williams2016incommensurate} and H$_3$LiIr$_2$O$_6$~\cite{kitagawa2018spin}, as well as $\alpha$-RuCl$_3$~\cite{plumb2014spin, sandilands2015scattering, sears2015magnetic, majumder2015anisotropic, johnson2015monoclinic, sandilands2016spin, banerjee2016proximate, sears2017phase, banerjee2017neutron, baek2017evidence, do2017majorana, banerjee2018excitations, hentrich2018unusual, kasahara2018majorana}.

\begin{figure}
\includegraphics[scale=0.95]
                {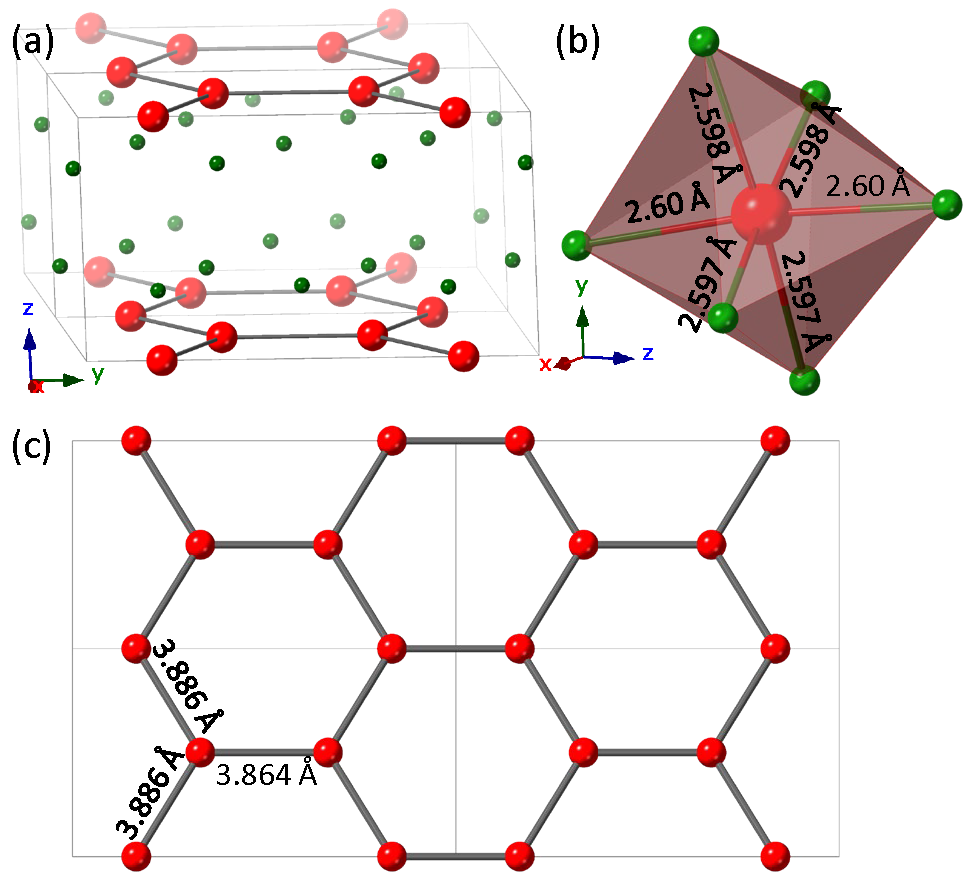}
\caption{
\label{fig: 1}
Monoclinic crystal structure of \YBCL{} with $a=6.7291(3)$\AA, $b=11.6141(9)$\AA, $c=6.3129(3)$\AA~and $\beta=110.5997(7)$ obtained at 10 K.  Refined structure parameters are further described in SI\cite{supp}.  (a) \YBCL{} structure consisting of alternating planes of Yb$^{3+}$ cations (red spheres) forming a honeycomb lattice in the $ab$ plane, with Cl$^{-}$ anions (green spheres) separating the layers. (b) The crystal field environment surrounding the rare earth ions consists of 6 Cl ions arranged in a distorted octahedron with $C_2$ point group symmetry.  (c) Single layer of Yb ions showing the honeycomb lattice arrangement in the monoclinic $ab$ plane with Yb-Yb distances at 10 K.}
\end{figure}

Recently, \YBCL{} has been proposed to be a candidate material for Kitaev physics on a honeycomb lattice \cite{ybcl3_ni,chen_theory}. \YBCL{} crystallizes in the monoclinic space group $C12/m1$ ($\#$12).  The crystal structure is composed of layers of Yb$^{3+}$ ions coordinated by slightly distorted Cl octahedra as illustrated in Fig.~\ref{fig: 1}. Despite being formally monoclinic at 10 K, the Yb-Yb distances of 3.864 \AA~and 3.886 \AA~and the Cl-Yb-Cl bond angles of 96.12$^{\circ}$ and 96.73$^{\circ}$ are nearly identical\cite{supp}.  The result of this atomic arrangement are well-separated, nearly-perfect honeycomb layers of Yb$^{3+}$ ions in the $ab$-plane as shown in Fig.~\ref{fig: 1}(a,c).   The environment surrounding the Yb$^{3+}$ cations depicted in Fig.~\ref{fig: 1}(b) consists of $6$ Cl$^-$ anions arranged in distorted octahedra where the $b$-axis is the unique $C_2$ axis. Xing, et al. \cite{ybcl3_ni} have reported that \YBCL{} undergoes short range magnetic ordering at 1.2 K.  A small peak in the heat capacity at $0.6$~K may indicate a transition to long range magnetic order.  Moreover, the field dependence of the inferred ordering temperature suggests that the interactions in \YBCL{} are 2-dimensional.  On the other hand, Yb-based quantum magnets have been the subject of several recent investigations and, surprisingly, in many cases these materials have been found to possess strong effective Heisenberg exchange interactions \cite{Rau_2016,haku2016,Park2016,Rau_2018,Wu2016,Wu2019,rau_yb}. Thus, key open questions for \YBCL{} are the nature of the spin Hamiltonian and the role of potential Kitaev terms.  It is likewise important to determine the single-ion ground state out of which the collective physics grows and additionally if the ground state doublet is well isolated and can be considered to be in the effective quantum spin-1/2 limit.  In this paper we address these issues using inelastic neutron scattering and thermodynamic measurements to study the crystal field and low energy excitation spectrum in polycrystalline samples of \YBCL{}.

Anhydrous beads of \YBCL{} and \LUCL{} were purchased from Alfa Aesar and utilized in the experimental work presented here.  Additional information and results of sample characterization are provided in the SI\cite{supp}.  Refinements of neutron powder diffraction data did not reveal any significant chlorine deficiency or secondary phases\cite{supp}.

The crystal field excitations were measured with inelastic neutron scattering performed with the SEQUOIA spectrometer at the Spallation Neutron Source at Oak Ridge National Laboratory (ORNL)~\cite{SEQref}. Approximately 4.2 g of polycrystalline \YBCL{} and 2.5 g of its non-magnetic equivalent \LUCL~were loaded into cylindrical Al cans and sealed under helium exchange gas. The use of the \LUCL{} measurement as a background subtraction is described in the SI\cite{supp}.   The samples and an equivalent empty can for Al background subtraction~\cite{aluminum} were measured at $T=5$~K, $95$~K and $185$~K, with incident energies, $E_i=6$~meV, $45$~meV and $60$~meV with the high resolution chopper.  Unless otherwise noted, all inelastic data presented here have had the measured backgrounds subtracted with the data reduced using the software packages Dave~\cite{dave} and MANTID~\cite{mantidplot}.

\begin{figure}
\includegraphics[scale=0.75]
                {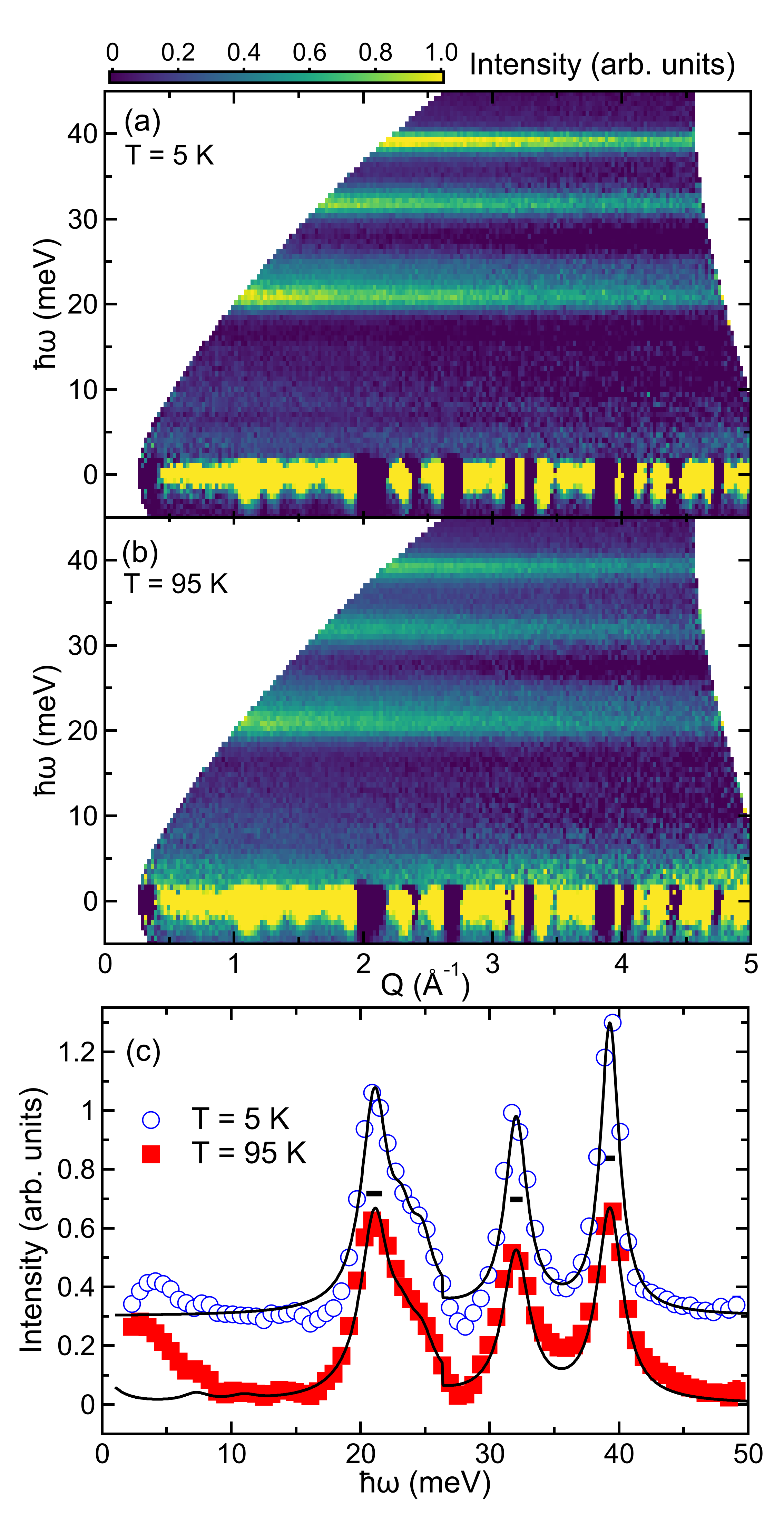}
\caption{
\label{fig: seqspectra}
Dynamic structure factor $S(|Q|,\hbar \omega)$ of \YBCL{} collected with SEQUOIA with
E$_i$=60 meV at (a)$T=5$~K and (b)$T=95$~K. The nonmagnetic background determined from \LUCL{} has been subtracted.   Crystal field excitations are visible at $\hbar \omega=$ $21$, $32$ and $39$ meV. (c) Comparison of the intensity of the crystal field transitions at $T=5$~K and $T=95$~K for \YBCL{}, in the momentum transfer range $Q=[0,3]$ \AA$^{-1}$. The solid lines are the results of the CEF analysis using Eq.~\ref{eq: 1}.  Horizontal black lines represent the instrumental resolution. The $T=5$~K data and refined model are offset by 0.3 units along the vertical axis.}
\end{figure}

Figures~\ref{fig: seqspectra}(a) and (b) show the inelastic neutron scattering spectra as a function of wave-vector transfer, $Q$, and energy transfer, $\hbar\omega$, measured at $T=5$~K and $95$~K respectively.   Figure~\ref{fig: seqspectra}(c) is the wave-vector integrated scattering intensity from the $E_i=60$~meV measurements for $Q<3$~\AA$^{-1}$.    The prominent higher energy modes are identified as crystal field excitations both from their $Q$-dependence and from comparison with the nonmagnetic analog \LUCL{}\cite{supp}. At $T=5$~K they are centered at energy transfers of $\hbar\omega=20.9$, $31.7$, and $39.5$~meV.  The 20.9 meV mode is noticeably broadened toward higher energy transfers.   Increasing temperature reduces intensity but does not appreciably shift or broaden these transitions, consistent with the behavior expected for crystal field excitations.  Note, that there are some lower energy acoustic phonon modes in the data that are not well subtracted, particularly near 4 meV.

\begin{figure}
\includegraphics[scale=0.9] {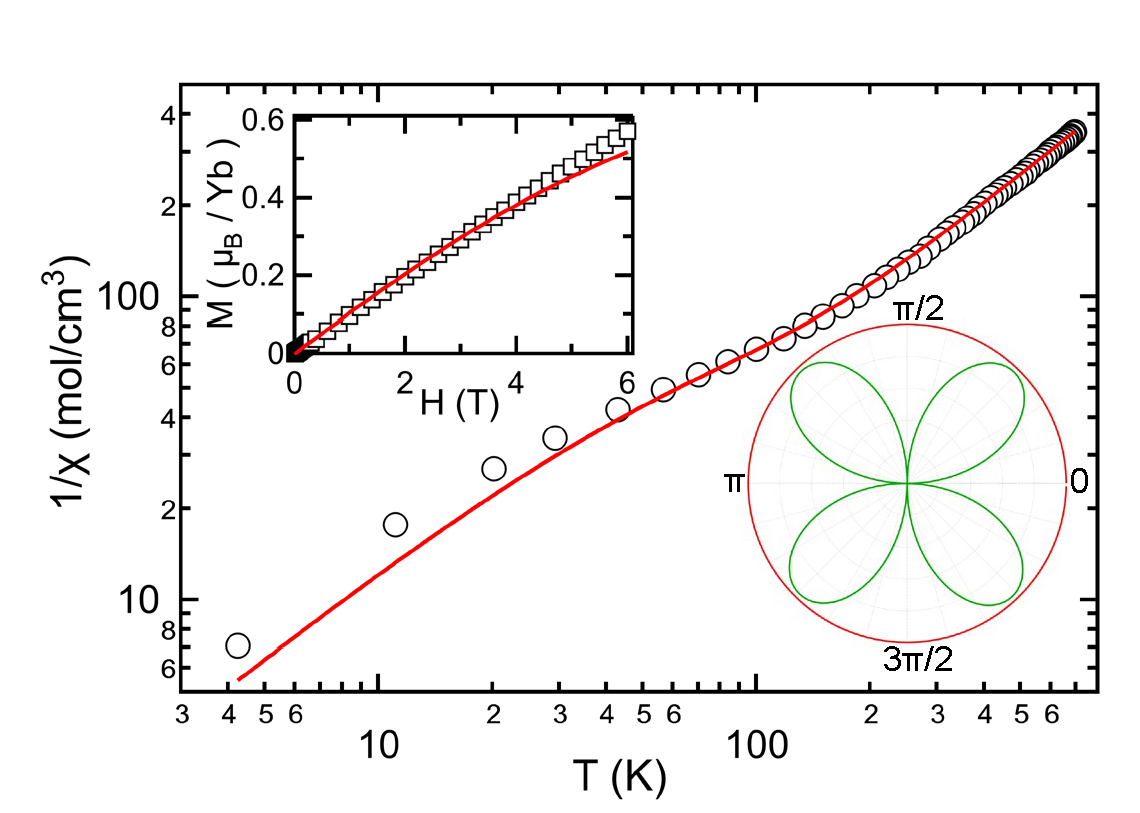}
\caption{
\label{fig: 2}
Top: Inverse magnetic susceptibility $\chi$ as a function of temperature for polycrystalline \YBCL{} in the range $4\leq$T$\leq 700$ K shown on a log-log scale for $H =  1 T$. The red line is the result of a simultaneous fit of the CEF model to the inelastic neutron scattering data (Fig.~\ref{fig: seqspectra}), the magnetic susceptibility, and the magnetization at 10 K. The top inset shows the calculated powder averaged magnetization at 10 K compared with the experimental data. The bottom inset shows the calculated torque diagram using the crystal field parameters (green curve) at $2.1$ K under an applied field of $5$ T (red circle) in the $ab$ plane as measured
in Ref. [\onlinecite{ybcl3_ni}]. }
\end{figure}

To understand the nature of the crystal field spectrum we analyze the energy levels following a formalism described by Wybourne~\cite{Wybourne,Judd1,Judd2} and Stevens~\cite{Stevens}. Given the $C_2$ site symmetry of the local Yb environment, the crystal field Hamiltonian consists of $14$ parameters~\cite{Walter}. Prather's convention~\cite{Prather} for the minimal number of crystal field parameters was achieved by rotating the environment by $\pi/2$ around the a-axis ($\hat{x}$-axis), i.e. the axis of quantization becomes the b-axis (the $\hat{z}$-axis in the rotated coordinate system).  To constrain the parameters, we simultaneously fit the neutron scattering data at 5 and 95 K (Fig.~\ref{fig: seqspectra}(c)), the static magnetic susceptibility between 4 and 700\,K (Fig.~\ref{fig: 2}) and the field-dependent magnetization at 10 K (inset of Fig.~\ref{fig: 2}).

Hund's rules state that, for a $4f^{13}$ ion, $L=3$ and $S=1/2$, thus $J = \| L + S\| = 7/2~$~\cite{Kittel}.  Therefore the crystal field Hamiltonian can be written in terms of Steven's operators as
\begin{eqnarray}
H = \sum_{n=2}^6 \sum_{m=0}^{\leq n} B_n^m \hat{O}_n^m + \sum_{n=4}^6 \sum_{m=2}^{\leq n} B(i)_n^m \hat{O}(i)_n^m
\label{eq: 1}
\end{eqnarray}
for $n$ even, where $B_n^m$ are the crystal field parameters, and $\hat{O}_n^m$ are the Steven's operators~\cite{Hutchings} both in real and imaginary ($i$) form.  Once Eq.~\ref{eq: 1} is diagonalized, the scattering function, $S(|Q|,\hbar \omega)$, can be written as
\begin{equation}
S(|Q|,\hbar\omega) = 
\sum_{i,i'}\frac{(\sum_{\alpha}  |\langle i {| J_{\alpha} | i'\rangle |}^2) \mathrm{e}^{-\beta E_{i}} }{\sum_j \mathrm{e}^{-\beta E_{j}}} L(\Delta E + \hbar \omega,\Gamma_{i,i'})
\label{eq: 2}
\end{equation}
where $\beta = 1/k_BT$, $\alpha = x,y,z$, $\Delta E = E_{i} - E_{i'}$, and $L(\Delta E + \hbar \omega, \Gamma_{i,i'})$ is a Lorentzian function\footnote{As explained in greater detail below, a constrained two component Lorentzian has been used for the crystal field level at 21 meV} with halfwidth $\Gamma_{i,i'}$ that parameterizes the lineshape of the transitions between CEF levels (eigenfunctions of Eq. (\ref{eq: 1})) $i \rightarrow i'$.  We calculate the scattering function using this formalism, compare these values with the experimental data, and then vary the crystal field parameters to minimize the $\chi^2$ difference between the model and the data shown in Figs.~\ref{fig: seqspectra}(c) and Fig.~\ref{fig: 2}.

\begin{table} 
\begin{tabular}{c|c|c|c|c}
\hline\hline
$B_2^0$ & $B_2^2$ & $B_4^0$ & $B_4^2$ & $B_4^4$ \\ 
$-2.820$ & $-25.956$ & $6.170$ & $42.336$ & $-36.335$ \\
\hline
$B_6^0$ & $B_6^2$ & $B_6^4$ & $B_6^6$ & \\
$-3.004$ & $10.764$ & $7.482$ & $49.327$ & \\
\hline
$B(i)_4^2$ & $B(i)_4^4$ & $B(i)_6^2$ & $B(i)_6^4$ & $B(i)_6^6$ \\
$-6.42\times10^{-3}$ & $-0.015$ & $8.56\times10^{-3}$ & $-0.067$ & $-0.036$ \\
\hline\hline
\end{tabular}
\caption{
\label{tab: 1}
Refined crystal field parameters in units of meV determined as described in the text.  Each coefficient is presented divided by the corresponding Steven's parameter $\alpha_J$, $\beta_J$ and $\gamma_J$~\cite{Hutchings}.}
\end{table}

The refinement of the Hamiltonian (Eq.~\ref{eq: 1}) in the scattering function described in Eq.~\ref{eq: 2} yields the crystal field parameters presented in Tab.~\ref{tab: 1} and the set of eigenfunctions written in Tab. II of the SI\cite{supp}. The ground state eigenfunction is found to be 
\begin{equation}
0.695 \ket{\pm \frac{7}{2}} -0.318\ket{\mp \frac{5}{2}} +0.546\ket{\pm \frac{3}{2}} -0.343\ket{\mp \frac{1}{2}}.
\label{gs}
\end{equation}
The imaginary part of the eigenfunction is not shown because it was determined to be $\approx2$ orders of magnitude smaller than the real part.  The calculated $S(|Q|,\hbar\omega)$ is plotted at both temperatures and is shown in Fig.~\ref{fig: seqspectra}(c) as solid lines.  The integrated intensity of the three crystal field excitations is reproduced as is the magnetic susceptibility (Fig. \ref{fig: 2}) and the field dependent magnetization at 10 K (inset of Fig. \ref{fig: 2}). 

The CEF model demonstrates that the Yb$^{3+}$ ions have a planar anisotropy and a calculated magnetic moment of $2.24(5)\mu_B$/ion is obtained for this ground state.   The calculated components of the $g$-tensor for the ground state doublet, using the convention described above, for \YBCL{} are $g_{z}=4.2(2)$, $g_x=3.8(2)$, and $g_y=2.2(2)$, which shows somewhat more anisotropy than Ref. [\onlinecite{ybcl3_ni}].  Additionally, using the crystal field model derived here as a starting point, we calculated a magnetic torque diagram at 2.1 K for an applied field of 5 T (Fig. \ref{fig: 2} inset).  The result reproduces the data  in Ref. [\onlinecite{ybcl3_ni}] (note the difference in coordinate conventions), demonstrating that the crystal field ground state is anisotropic independent of any additional exchange anisotropy.

\begin{figure}
\includegraphics[scale=0.9]
{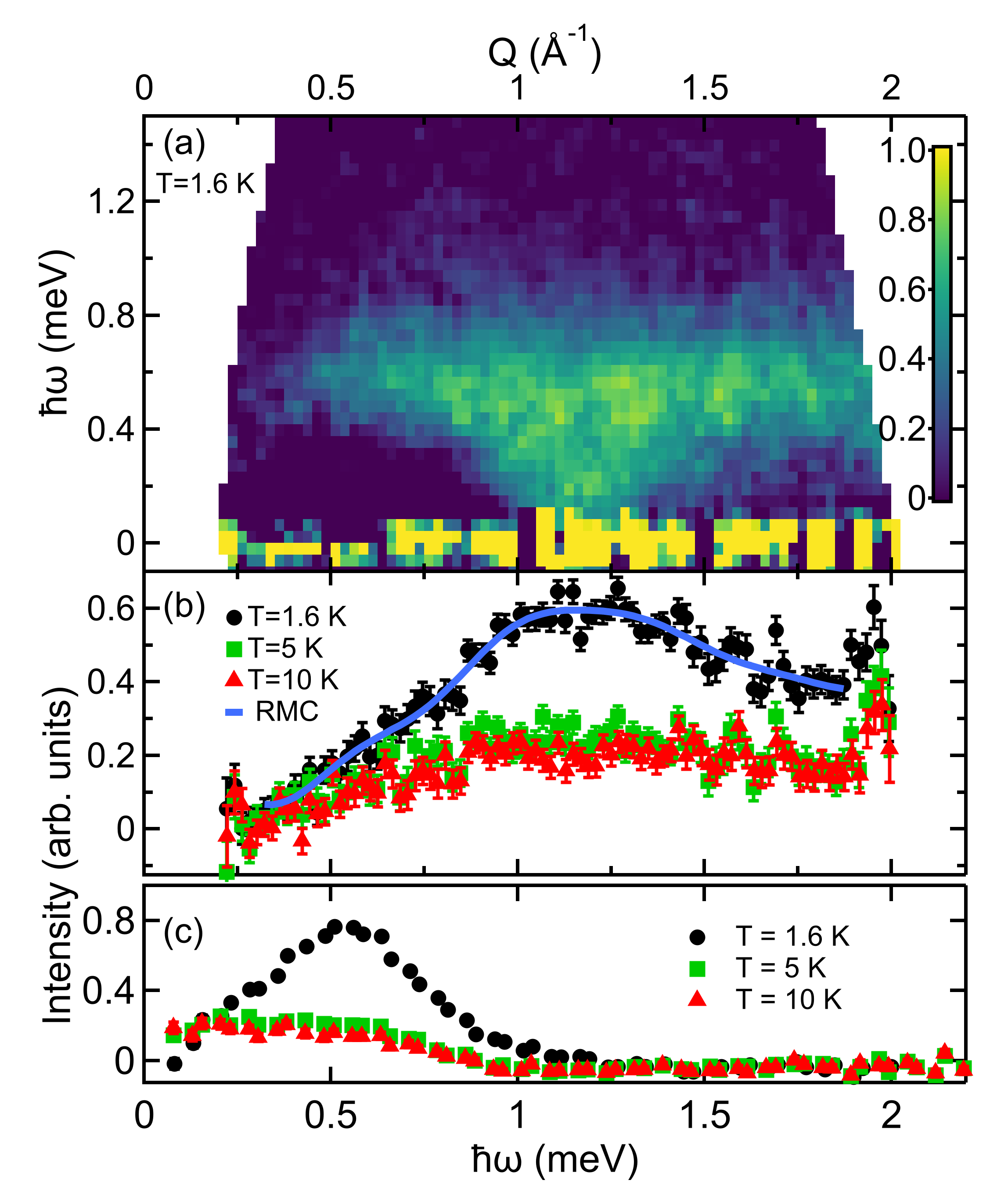}
\caption{
\label{fig: 6}
Low energy magnetic spectrum of \YBCL.  All data have had the $T=100$~K \YBCL{} measurement subtracted as a background. (a) Scattering intensity as a function of $Q$ (top axis) and $\hbar\omega$.  (b) Scattering intensity as a function of $Q$ (top axis) integrated over $\hbar\omega=[0.1,1.2]$~meV.  The solid line is the RMC calculation described in the text. (c) Scattering intensity as a function of $\hbar\omega$ (bottom axis) integrated over $Q=[0,2]$ \AA$^{-1}$.}
\end{figure}

Despite the overall quality of the fits, one aspect of the crystal field excitation spectrum remains puzzling.  The lineshape of the crystal field excitation centered at 21 meV extends toward higher energies. A similar broadening is not observed for the other crystal field excitations at 32 and 40 meV. Thus the broadening is a characteristic of the level at 21 meV and not of the ground state.   To fully account for the spectral weight, we have modeled the lineshape for this excitation as two constrained Lorentzians with the widths fixed to be the same and the positions offset by a fixed amount. The lack of observable impurity peaks in the neutron diffraction data\cite{supp} suggests that this effect is not due to an impurity phase. Deviations from ideal Cl stoichiometry are similarly hard to detect.  Another possibility is that stacking faults result in a variation of the crystal field potential along the c-axis, though this was not evident in the diffraction data.  The level at 21 meV would be more strongly affected by such stacking faults given the strong charge density out of the plane for this eigenfunction (see SI\cite{supp} Fig. S4 for plots of the charge density for each eigenfunction). Additionally, first principles calculations of the phonon density of states suggests that this feature is not the result of hybridization of the crystal field level with a phonon\cite{supp}. Careful studies of single crystals are required to further understand the origin of this broadening.  Finally, we note that using a single Lorentzian in the CEF modeling does not significantly change the refined CEF parameters as the additional spectral weight is relatively small.

To probe for low-energy magnetic correlations, we performed inelastic neutron scattering measurements using the HYSPEC instrument~\cite{hyspec} at ORNL.  For these measurements, the same sample used in the SEQUOIA measurements was cooled to $T=1.6$, $5$, and $10$~K and  measured with $E_i=3.8$~meV at two different positions of the detector bank to cover a large range of $Q$.  A $T=100$~K measurement of the \YBCL{} sample was used as the background to isolate the magnetic scattering.  Figure~\ref{fig: 6}(a) shows the energy and wave-vector dependent magnetic spectrum in \YBCL{}. A broad dispersive mode with additional scattering is evident. The additional scattering may be due to a quantum continuum however, other explanations such as broadened excitations from a short ranged ordered state, magnon decay, etc, cannot be excluded with the data at hand.  The $Q$ integrated scattering intensity in Fig.~\ref{fig: 6}(c) shows a single peak at approximately 0.5 meV with no indication of a spin gap or additional scattering intensity above approximately 1.3 meV energy transfer.  Given that long range magnetic order occurs at a maximum temperature of 0.6 K\cite{ybcl3_ni}, the energy scale of the spin excitations suggests low dimensional and/or frustrated spin interactions in \YBCL{}. The $\hbar\omega$ integrated intensity in Fig.~\ref{fig: 6}(b) is a broad function which peaks at approximately $Q=1.1$~\AA$^{-1}$ likely corresponding to the reciprocal lattice points (1 1 0) and (0 2 0), which is consistent with spin correlations within the basal plane. The data in Fig.~\ref{fig: 6}(c) were collected at $T$=1.6~K, which is at lower $T$ than the maximum in the specific heat capacity\cite{ybcl3_ni,supp}.  Thus, the low temperature spin excitations may be responsible for a portion of the loss of entropy despite the lack of apparent long range order.  Additional measurements using single crystals are required to fully understand the nature of the magnetic ground state and the spin excitation spectrum.

To investigate the spin-spin correlations in \YBCL{}, we performed Reverse Monte Carlo (RMC) calculations as implemented in Spinvert~\cite{Paddison2013} (see SI\cite{supp} for more details). Within this approximation, we fit the integrated intensity of the low energy excitation spectrum as a function of $Q$. Ising, XY, and Heisenberg types of spin correlations were all tried as initial starting points for the simulations (See the SI\cite{supp} for further detail). The result of the RMC modeling is shown as a solid line in Fig.~\ref{fig: 6}(b). The radial spin-spin correlation function was calculated for each final spin configuration as a means to investigate the orientation of the spins with respect to each other. Assuming a purely hexagonal geometry, nearest neighbor spins are antiferromagnetically correlated, second neighbor spins have weak ferromagnetic correlations, while there is a rapid decay of spin correlations at larger distances. This result is independent of the type of starting correlation used for the modeling.

We analyzed the spectroscopic properties of the interesting quantum magnet \YBCL. Our studies show that \YBCL{} has crystal field excitations at $\hbar\omega=20.9$, $31.7$, and $39.5$~meV.  The ground state is a well separated effective spin-1/2 doublet with easy plane ansisotropy and an average magnetic moment of $2.24(5)\mu_B$/ion.  At $T$=1.6~K, where long range order is not believed to exist, the low energy dynamics of the \YBCL{} are consistent with a low dimensional interacting spin system with antiferromagnetic nearest neighbor correlations.

\begin{acknowledgments}
We thank J. Liu for useful discussions.  This work was supported by the U.S. Department of Energy (DOE), Office of Science, Basic Energy Sciences, Materials Sciences and Engineering Division. This research used resources at the High Flux Isotope Reactor and Spallation Neutron Source, DOE Office of Science User Facilities operated by the Oak Ridge National Laboratory. The work of G.B.H. at ORNL was supported by Laboratory Director’s Research and Development funds. The computing resources for VASP and INS simulations were made available through the VirtuES and the ICE-MAN projects, funded by Laboratory Directed Research and Development program, as well as the Compute and Data Environment for Science (CADES) at ORNL.
\end{acknowledgments}

\renewcommand{\theequation}{S\arabic{equation}}
\renewcommand{\thefigure}{S\arabic{figure}}
\renewcommand{\thetable}{S\arabic{table}}

\setcounter{figure}{0}
\bigskip
\bigskip

\centerline{\textsc{\textbf{\LARGE Supplemental Information}}}

\section{Sample Preparation}

Anhydrous beads of \YBCL{} (Alfa Aesar stock 40653) were utilized and care was taken to avoid exposure to air.  The material was received in an argon-filled glass ampoule, which was opened inside a helium-filled glovebox.  The material was stored in the glovebox prior to loading aluminum canisters for neutron scattering measurements.  An indium seal was tightened inside the helium glovebox and checked with a helium leak detector.  A powder sample of anhydrous \LUCL{} (Alfa Aesar stock 35802) was handled in the same way so that a non-magnetic analogue could be examined.

\begin{figure}[h]
\includegraphics[width=0.99\columnwidth]
                {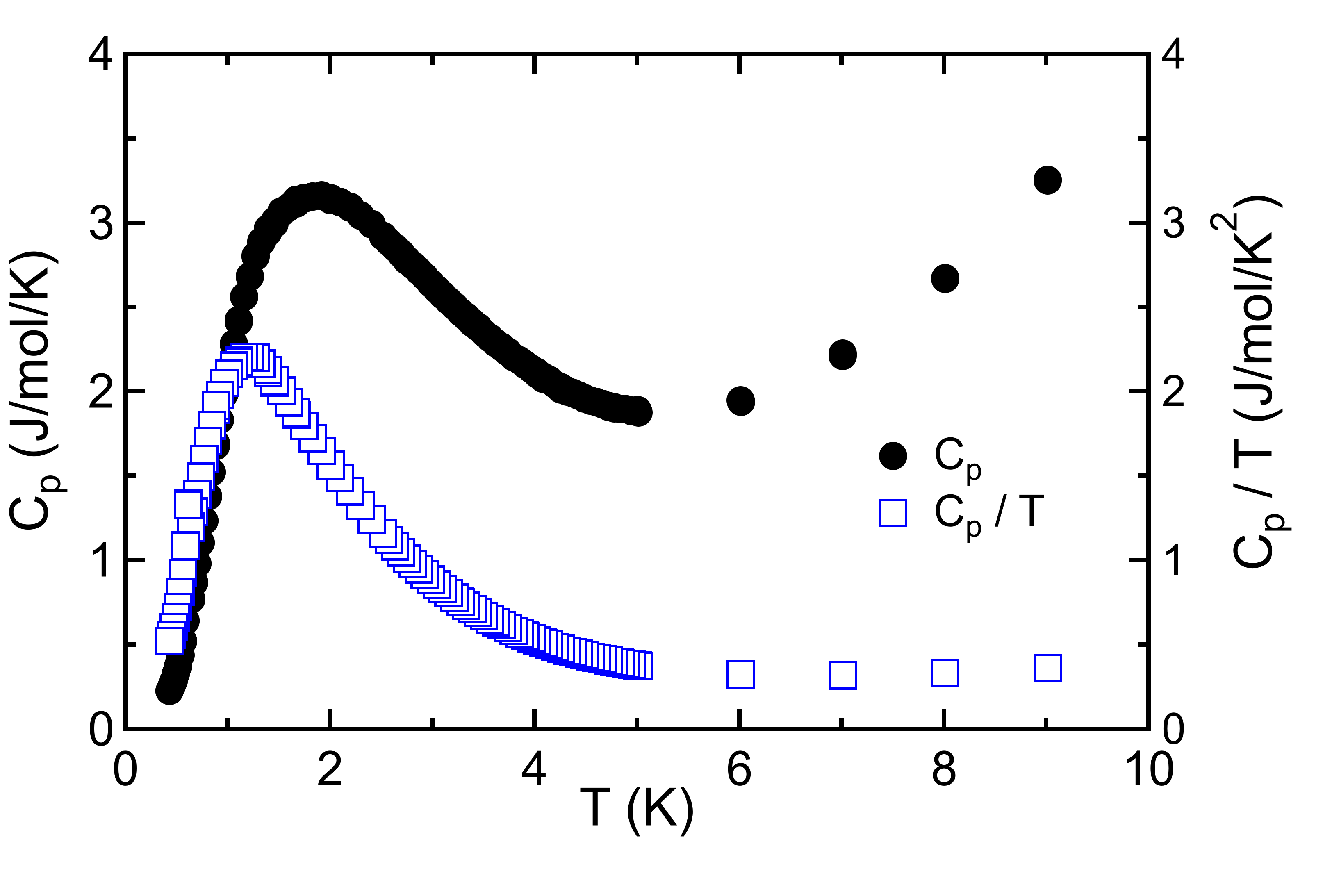}
\caption{
\label{specheat}
Specific eat data for polycrystalline samples of \YBCL{}.  The scale is the same for $C_p$ and $C_p/T$.  These results are consistent with the single crystals results of Ref. [\onlinecite{ybcl3_ni}].}
\end{figure}

\begin{figure}[ht]
\includegraphics[width=0.9\columnwidth]
                {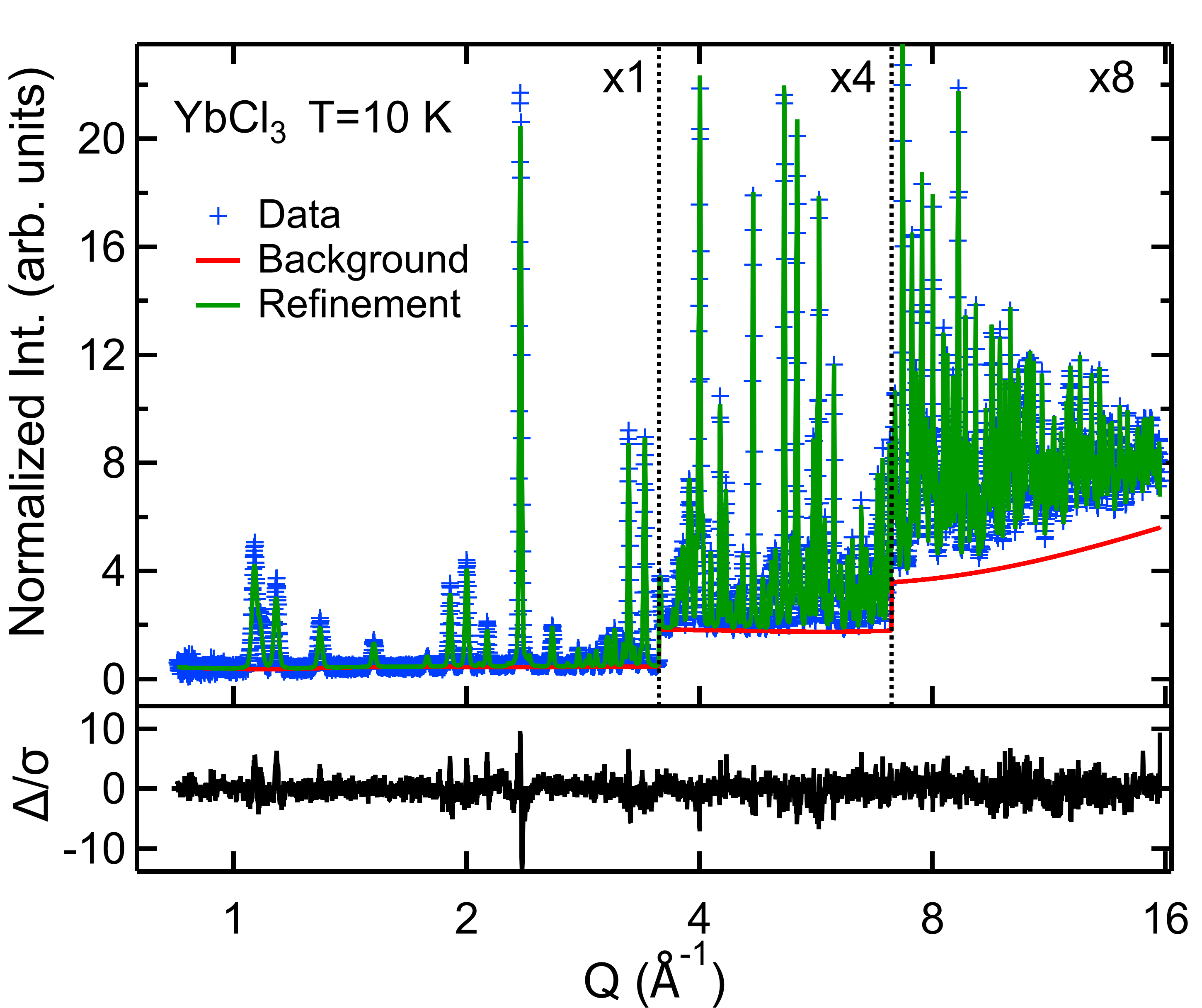}
\caption{
\label{powgen}
Neutron diffraction data and Rietveld refinement of the $C12/m1$ structural model for \YBCL{}.  The multiplying factors are used to provide additional detail. The bottom panel shows the difference between our refinement and the data.}
\end{figure}

\section{Heat Capacity}
Specific heat measurements down to 0.4 K were performed in a Quantum Design Physical Property Measurement System equipped with a $^3$He insert. The low-temperature specific heat $C_P(T)$ (left axis) and $C_P(T)/T$ (right axis) of \YBCL{} are shown in Fig. \ref{specheat}. Our $C_P(T)$ measurement shows a broad feature centered around 1.9 K, which corresponds to a maximum near 1.2 K in $C_P(T)/T$.  Upon close inspection, a very weak feature can also be inferred near 0.6 K.  These results are consistent with those reported by Xing et. al.[\onlinecite{ybcl3_ni}].  As demonstrated by Xing et. al., the majority of the magnetic entropy is lost above 0.6 K.

\begin{table*}
\begin{tabular}{ccccc|cccc}
\hline \hline
                           & \multicolumn{4}{c|}{\begin{tabular}[c]{@{}c@{}}\\ T=10 K \quad $R_w$=2.42\\ \\ \quad a=6.7291(3) \AA,~b=11.6141(3) \AA,~c=6.3121(3) \AA~\quad \\ 
                           $\alpha=\gamma=90^\circ$, $\beta=110.5997(7)^\circ$ \\ \\ \end{tabular}} & \multicolumn{4}{c}{\begin{tabular}[c]{@{}c@{}}\\ T=100 K \quad $R_w$=2.93\\ \\ \quad a=6.7326(5) \AA,~b=11.6201(1) \AA,~c=6.3280(5) \AA~\quad \\ $\alpha=\gamma=90^\circ$, $\beta=110.551(1)^\circ$ \\ \\ \end{tabular}} \\ \hline \hline
\multicolumn{1}{c|}{Atom}  & x                   & y                   & z                   & Occ                   & x                   & y                   & z                   & Occ                    \\ \hline
\multicolumn{1}{c|}{Yb}    & 0                   & 0.1663(6)                   & 0                   & 1.0                     & 0                   & 0.1665(1)                   & 0                   & 1.0                      \\
\multicolumn{1}{c|}{Cl(1)} & \quad 0.2589(1) & 0.3214(1) & 0.2431(1) & 0.992(3)                  & \quad 0.2587(1) & 0.3209(1) & 0.2422(1) & 0.997(4)                      \\
\multicolumn{1}{c|}{Cl(2)} & \quad 0.2188(1) & 0 & 0.2493(1) & 0.997(4)                  & \quad 0.2179(2) & 0 & 0.2487(3) & 0.991(6)                   \\ \hline \hline
\end{tabular}
\caption{
\label{tab: 0}
Summary of Rietveld refinement parameters for \YBCL.  Data were refined in the space group $C2/m$ (12).}
\end{table*}

\section{POWGEN refinement results}

Neutron powder diffraction measurements were performed using the POWGEN time-of-flight diffractometer at the Spallation Neutron Source.~\cite{powgen2011}  Powder samples were measured in vanadium sample cans at $T=100$~K and $T=10$~K.  The data were corrected for neutron absorption.    The structural model was refined using the GSAS-II\cite{gsas2} package.  Data and refinement at 10 K are shown in Fig.~\ref{powgen}.   The refinement confirmed that our sample crystallizes into the monoclinic space group with $a=6.7291(3)$ \AA, $b=11.6141(1)$ \AA, $c=6.3120(3)$ \AA~and $\beta=110.5997(7)^\circ$. Alternately, as shown in the next section, neglecting the small monoclinic distortion, the in plance lattice parameters can be written in hexagonal notation as $a=b=6.720$ \AA~with $\gamma=120^\circ$.  Table~\ref{tab: 0} shows a summary of our refinement parameters at 100 K and 10 K. 


\section{Transformation equations from Monoclinic to Hexagonal Notation}

When the structure is weakly monoclinic it can be useful to transform the crystallographic parameters from monoclinic symmetry to hexagonal symmetry. Here we consider the transformation of a single layer from monoclinic to hexagonal notation.  If we keep the $\vec{a}$ constant for the two systems, then the relations connecting the two symmetries are:

\begin{eqnarray}
    \left\{
        \begin{array}{ll}
            \vec{a}_{mon} = \vec{a}_{hex}\\
            \vec{b}_{mon} = \vec{a}_{hex} + 2\vec{b}_{hex}\\
        \end{array}
    \right.
\end{eqnarray}

or equivalently:

\begin{eqnarray}
    \left\{
        \begin{array}{ll}
            \vec{a}_{hex} = \vec{a}_{mon}\\
            \vec{b}_{hex} = (\vec{b}_{mon} - \vec{a}_{hex})/2 \\
        \end{array}
    \right.
\end{eqnarray}

where the subscripts ``$mon$" and ``$hex$" denote monoclinic and hexagonal notation respectively. From these vectors the reciprocal lattice vectors can be calculated using the standard formulas\cite{Kittel}.

Therefore, in analogy, a monoclinic reflection $q=(h,k)$ can be transformed into its hexagonal equivalent as follows:

\begin{eqnarray}
 \begin{pmatrix}
 h \\
 k \\
 \end{pmatrix}_{mon}
=
 \begin{pmatrix}
 1 & 0  \\
 1 & 2  \\
 \end{pmatrix}
 \begin{pmatrix}
 h \\
 k \\
 \end{pmatrix}_{hex}
\end{eqnarray}

\begin{eqnarray}
 \begin{pmatrix}
 h \\
 k \\
 \end{pmatrix}_{hex}
=
 \begin{pmatrix}
 1 & 0  \\
 -1/2 & 1/2  \\
 \end{pmatrix}
 \begin{pmatrix}
 h \\
 k \\
 \end{pmatrix}_{mon}
\end{eqnarray}


\section{Sequoia crystal field background analysis}

\begin{figure*}
\includegraphics[width=1.75\columnwidth]
                {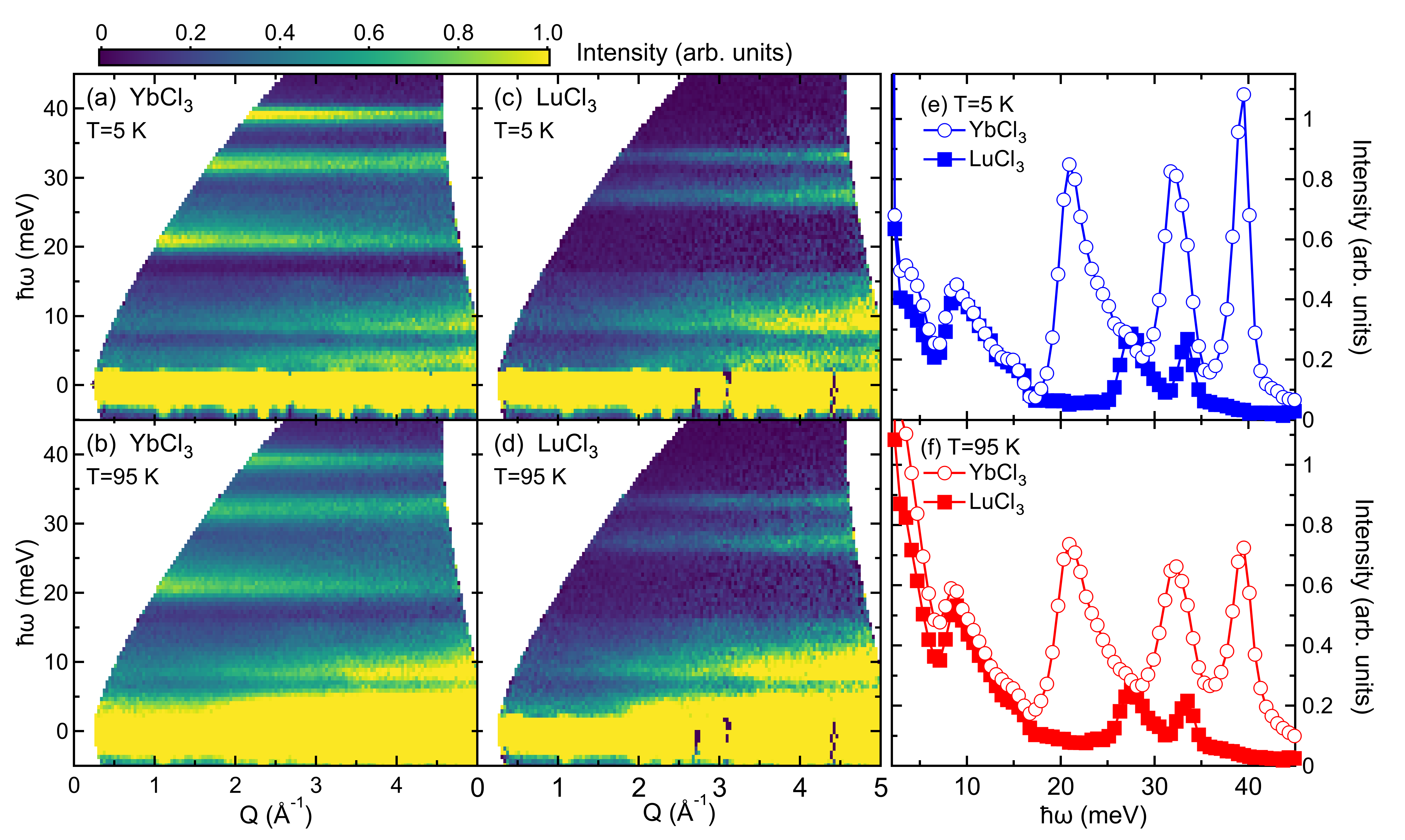}
\caption{
\label{ins_sup}
(a-d) Comparison of \YBCL~and \LUCL~data collected at SEQUOIA with E$_i$= 60 meV at 5 K and 95 K; an aluminum empty can data set has been subtracted to eliminate the contribution of the holder. (e-f) Cuts taken in the range Q=[0,3] \AA$^{-1}$
showing the comparison between the intensity of the crystal field transitions in \YBCL~and the phonon density of states in \LUCL. No phonons are evident
in the energy range $16\leq \hbar \omega \leq 25$ meV excluding a possible hybridization with the first crystal field level of \YBCL.}
\end{figure*}

In order to study the crystalline electric field (CEF) of \YBCL{} we first investigated the background of our compound in the proximity of the three crystal field
excitations. In principle, a good background subtraction to eliminate the phonon contamination can be done by measuring a non-magnetic equivalent
of the main sample under the same experimental conditions. \LUCL{} was used for this purpose.

We present in Fig.~\ref{ins_sup}(a-d) a comparison of the data set collected with SEQUOIA for \YBCL{} and \LUCL{} with E$_i$ = 60 meV at 5 K and 95 K.
An empty can data set at the same temperature has been subtracted to eliminate the contribution due to the sample holder. The last column of Fig.~\ref{ins_sup}(e-f)
shows cuts in the momentum transfer range Q=[0,3] \AA$^{-1}$ at both temperatures in analogy with our analysis in the main text, to compare the
phonon density of states and the crystal field spectrum.

The energy transfer range below 15 meV is dominated by the phonons of Yb(Lu), in particular the acoustic phonon 
at Q=4 \AA$^{-1}$ and the optical phonon at 9.8 meV have been used to normalize the intensities of the two data sets. Cuts were taken through the acoustic phonon away from the nuclear Bragg peak for both samples, and the integrated intensities were compared to find the proper normalizing factor. Then this quantity was cross-checked by taking cuts along the optic phonon at $Q\geq4$~\AA$^{-1}$, where the magnetic form
factor of Yb$^{3+}$ is small. With the data sets normalized in this manner, the \LUCL{} data set was then subtracted from the \YBCL{} data set. The resulting data sets were used in the refinements of the crystal field model.

The wave functions of the refined crystal field model of \YBCL{} are presented in Tab.~\ref{tab: 2}. The complex parts are omitted because they are approximately two orders of magnitude smaller than the real ones. One of the possible ways to visualize the
eigenfunctions is to calculate and plot the electronic charge density $\rho_e(\vec{r})=\langle \psi^\dagger | \psi \rangle$ of each of the crystal field levels.  In general,
under the influence of an external electric field, the crystal field states $|\psi_i \rangle$ can be written as:

\begin{eqnarray}
|\psi_i \rangle = \sum_m a_{im}| J, \mu + mp \rangle
\end{eqnarray}
where $\mu$ denotes the CEF quantum number, $m$ is an integer and $J$ is the total angular momentum value. Here we choose, the axis quantization, $\hat{z}$, to be parallel to the $b$-axis, then
following the procedure described in Ref.~\cite{Walter}, $\rho_e$ can be expressed in terms of spherical harmonics as:

\begin{eqnarray}
\rho_e(\vec{r}) = \sum_{k=0}^{min(2l,2J)} \sum_{q=-k}^{+k} \rho_{kq}(\vec{r})Y_k^q(\Omega)
\end{eqnarray}
where $l=3$ for rare earth ions, $Y_k^q(\omega)$ is a spherical harmonic, and $k$ is limited to even numbers due to the time reversal invariance of the charge density.

\begin{figure*}
\includegraphics[width=2.0\columnwidth]
                {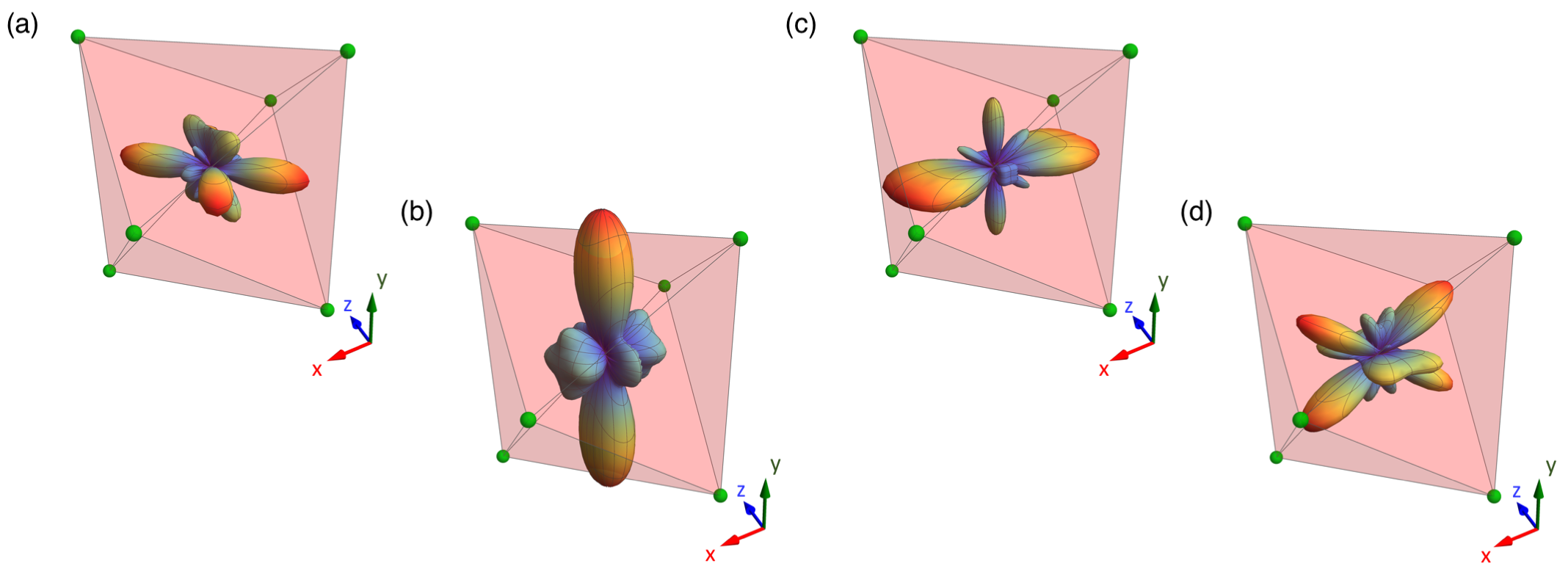}
\caption{
\label{fig: 3}
Real space representation of the angular part of the charge density $\rho_e(\vec{r})$ generated by the four crystal field levels. (a)-(d) correspond to the ground state (a) and then to progressively higher lying levels (b)-(d).  $R_{nl}^2(\vec{r})=1$ was used for ease of visualization.}
\end{figure*}

\begin{table*} 
\begin{tabular}{c|cccccccc}
\hline\hline
$m_J$ & $0.0$ & $0.0$ & $21.08$ & $21.08$ & $32.03$ & $32.03$ & $39.28$ & $39.28$ \\
\hline
$-7/2$ & & -0.695 & -0.196 & & & 0.689 & -0.062 & \\
$-5/2$ & -0.318 & & & -0.249 & -0.169 & & & 0.899 \\
$-3/2$ & & -0.546 & -0.428 & & & -0.691 & -0.205 & \\
$-1/2$ & -0.343 & & & -0.846 & -0.139 & & & -0.383 \\
$1/2$ & & 0.343 & -0.846 & & & 0.139 & 0.383 & \\
$3/2$ & 0.546 & & & -0.428 & 0.691 & & & 0.205 \\
$5/2$ & & 0.318 & -0.249 & & & 0.169 & -0.899 & \\
$7/2$ & 0.695 & & & -0.196 & -0.689 & & & 0.062 \\
\hline \hline
\end{tabular}
\caption{
\label{tab: 2}
Tabulated wave functions of the crystal field states in \YBCL{} obtained within the LS-coupling approximation. The crystal-field energies (in meV) are tabulated
horizontally, the $m_J$-values of the ground-state multiplet vertically. Only coefficients of the wave functions $> 10^{-3}$ are shown. The imaginary parts of the eigenvectors are omitted because they are $\approx2$ order of magnitude smaller
than the real ones.}
\end{table*}

Now, an eigenfunction can always be partitioned into a radial part $R_{nl}(\vec{r})$ which depends only on the quantum number $n$ and $l$, and 
an angular part that is a linear combination of spherical harmonics:

\begin{eqnarray}
\rho_e(\vec{r}) = R_{nl}^2(\vec{r}) \sum_{k=0}^{min(2l,2J)} \sum_{q=-k}^{+k} c_{kq}Y_k^q(\Omega)
\end{eqnarray}
or alternatively using the tesseral harmonics $Z_{kq}$ as~\cite{Hutchings}:

\begin{eqnarray}
\rho_e(\vec{r}) = R_{nl}^2(\vec{r}) \sum_{k=0}^{min(2l,2J)} \sum_{q=0}^{+k} \zeta_{kq}Z_k^q(\Omega).
\label{eq: 9}
\end{eqnarray}
Note that if a 2-fold axis perpendicular to the $p$-fold axis exists, then all the terms in the expansion are real.

The coefficients $\zeta_{kq}$ in Eq.~\ref{eq: 9} can be calculated using Steven's operators as:

\begin{eqnarray}
\zeta_{kq} = \sqrt{\frac{2k+1}{4\pi}}\theta_k b_{kq} \langle \psi^\dagger | O_k^q | \psi \rangle
\label{eq: 10}
\end{eqnarray}
where $\theta_k$ and $b_{kq}$ values can be found in Refs.~\cite{Hutchings,Walter1986} respectively, and $O_k^q$ are the Steven's operators.
It can also be verified that:

\begin{eqnarray}
\langle \psi^\dagger | O_k^q | \psi \rangle = 0
\label{eq: 11}
\end{eqnarray}
for $q$ not multiple of $p$ and:

\begin{eqnarray}
\langle \psi^\dagger | O_k^q | \psi \rangle = \sum_m a_m^2 \langle J, \mu+mp |O_k | J, \mu +mp \rangle
\label{eq: 12}
\end{eqnarray}
for $q$ multiple of $p$ and with m integer.

Finally, we can express the charge density using Eqs.~\ref{eq: 9}-\ref{eq: 12} as:

\begin{eqnarray}
\rho_e(\vec{r})= R_{nl}^2(\vec{r}) \sum_{k=0}^{min(2l,2J)}\sqrt{\frac{2K+1}{4\pi}}\theta_k 
\nonumber
\\
\sum_{q=mp}^k b_{kq}\langle \psi^\dagger | O_k^q | \psi \rangle Z_{kq}(\Omega)
\end{eqnarray}
for even $k$. The calculated angular part of the charge density for the 4 crystal levels is presented in Fig.~\ref{fig: 3}.
Note that we assumed $R_{nl}(\vec{r})=1$ for ease of visualization.

\begin{figure*}
\includegraphics[width=1.85\columnwidth]
                {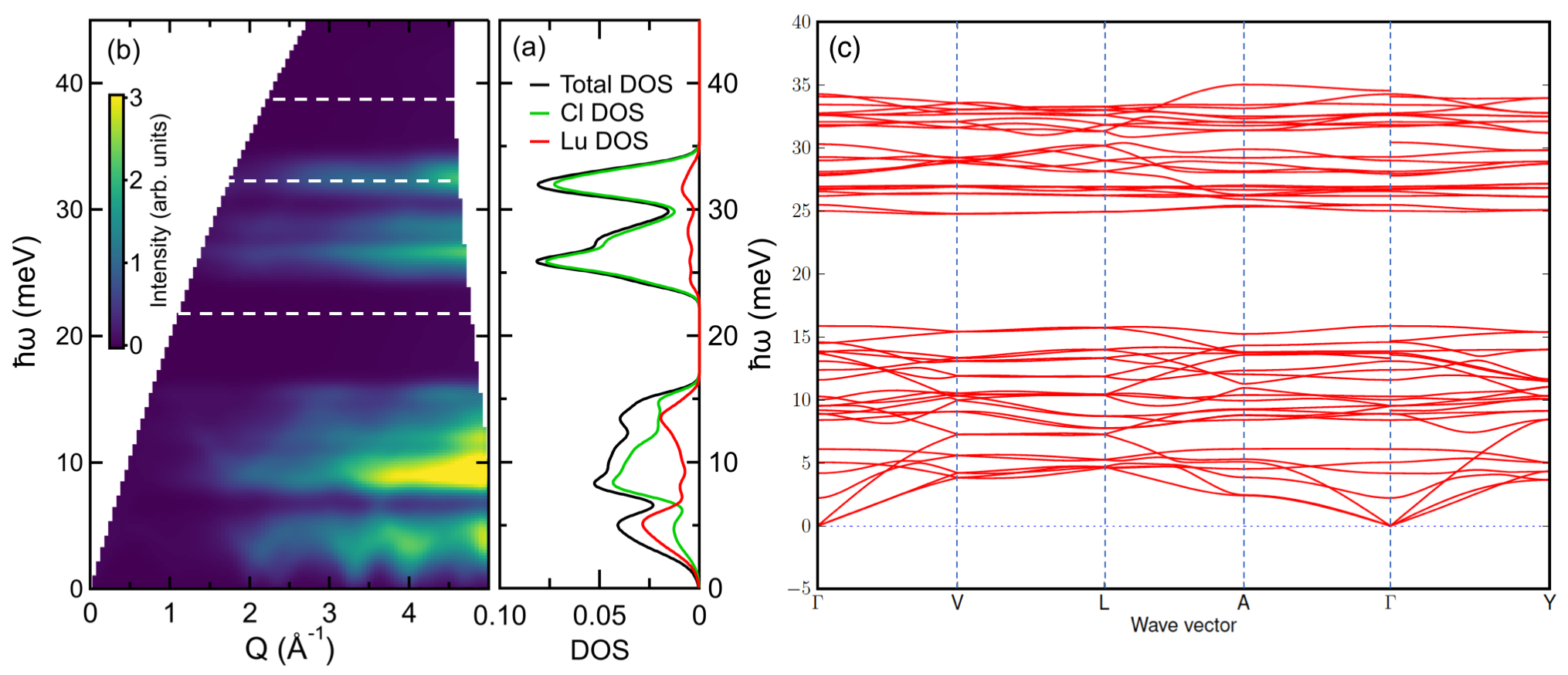}
\caption{
\label{fig: 4}
(a-b) DFT calculation of the phonon DOS for \LUCL{} as measured at the SEQUOIA spectrometer with E$_i=60$ meV. The calculations are in very good agreement with the data  
displayed in Fig.~\ref{ins_sup}(c); the three dashed lines are guide to eyes that show
the position of the CEF transitions of \YBCL. (c) DFT Calculated phonon dispersion across the highly symmetric directions of the Brillouin zones for \LUCL.}
\end{figure*}

\begin{figure*}
\includegraphics[width=1.72\columnwidth]
                {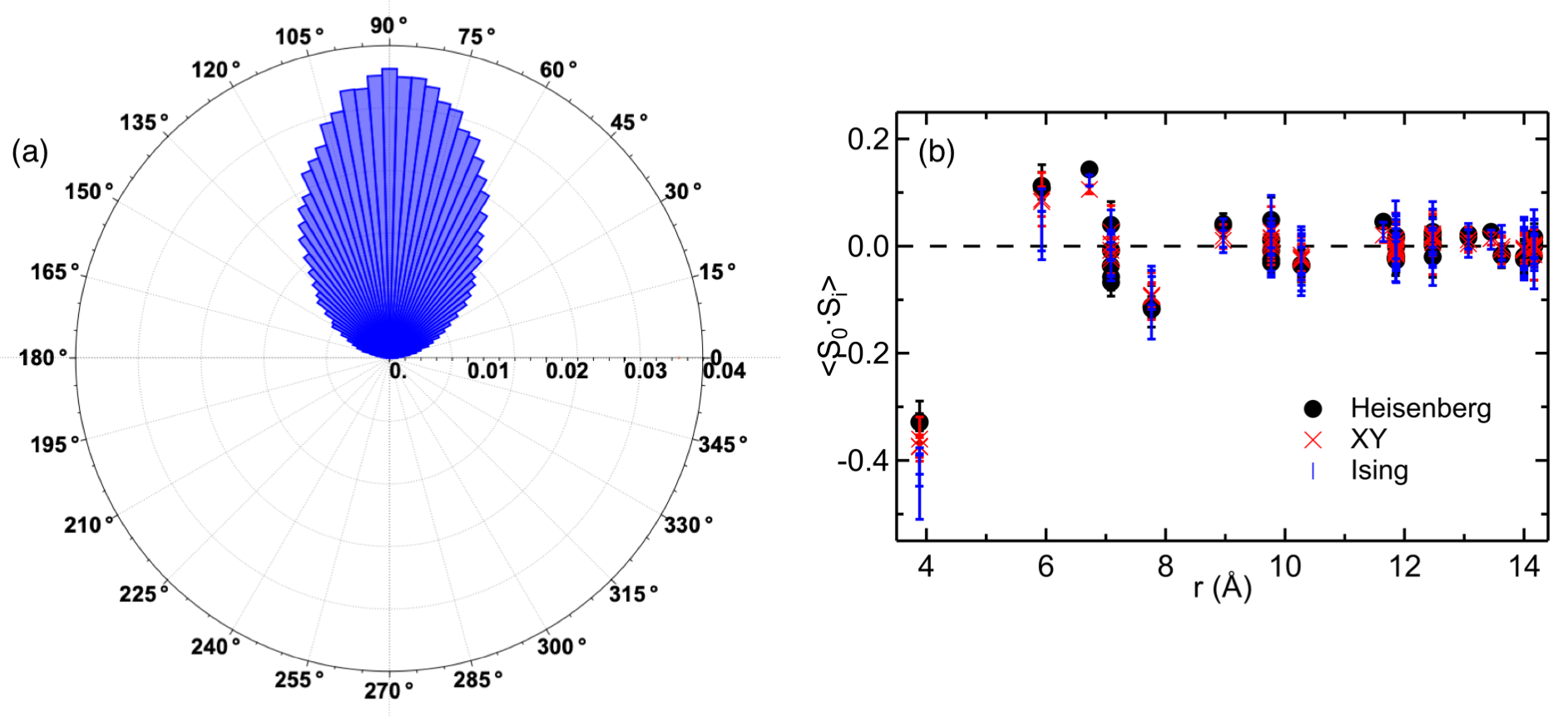}
\caption{
\label{fig: 7}
(a) The distribution of the $\theta$ angles within the range [0,$\pi$] calculated from the $\hat{z}$ axis (located at zero degree), shows that the spins are mainly laying in the $ab$ plane, as suggested by the anisotropy of the CF ground state. (b) Comparison of the calculated radial spin-spin correlation function for the three models used in the RMC simulation. The histogram shows that the there are antiferromagnetic nearest neighbor correlations. The presence of more than one marker at the same distance is due to slightly different bond-lengths 
of the spins in the lattice.}
\end{figure*}

\subsection{DFT Phonon Calculation on \LUCL}

\begin{figure*}
\includegraphics[width=1.72\columnwidth]
                {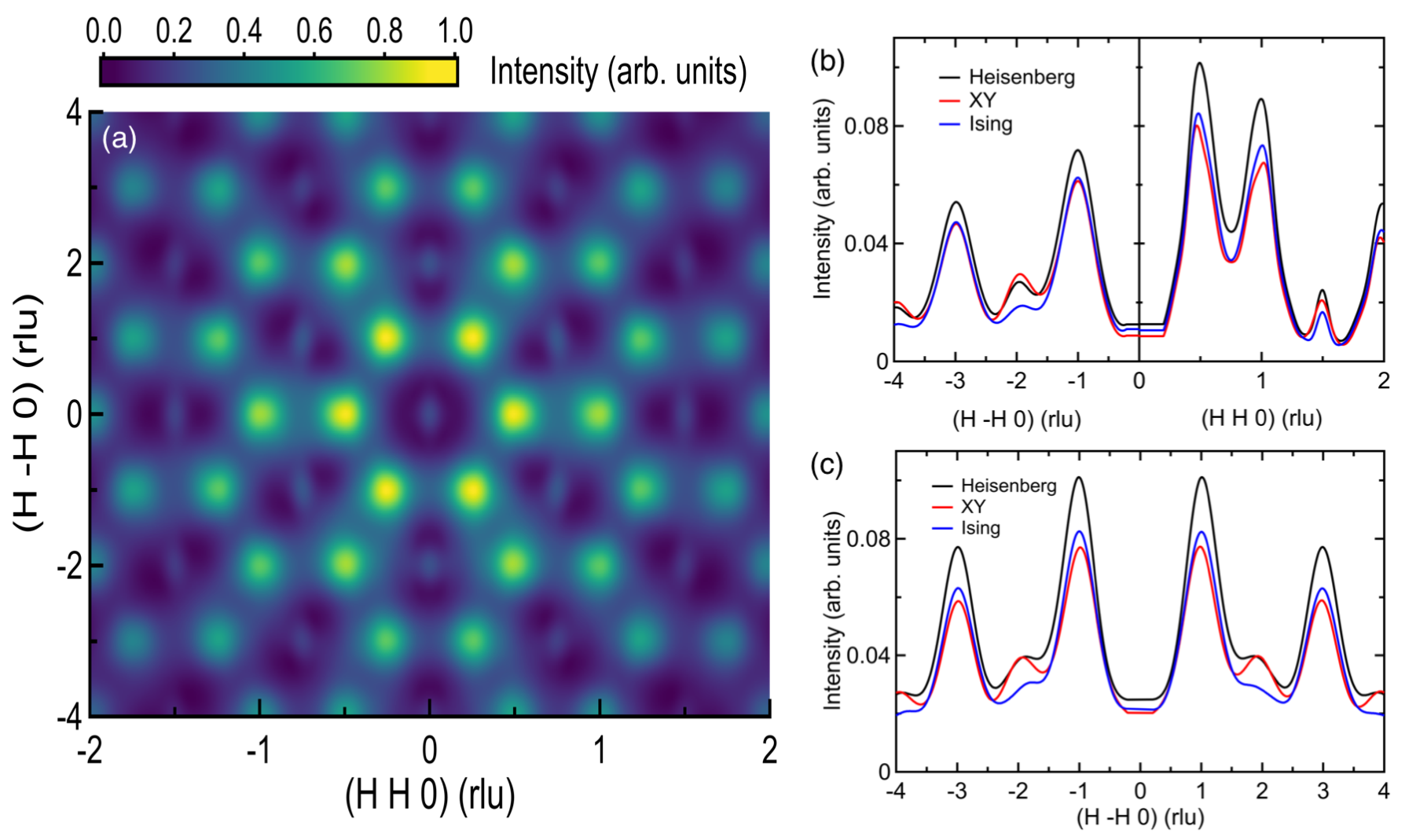}
\caption{
\label{fig: 5}
(a) Prediction of the diffuse scattering pattern for a single crystal in the (HH0) plane calculated using the best fitting parameters for the Heisenberg model in hexagonal coordinates. (b) Integrated intensity cuts along H-H0=[-0.25,0.25] r.l.u, and HH0=[-0.25,0.25] r.l.u,  
(c) Integrated intensity cuts along HH0=[0.16,0.36] r.l.u, showing the comparison of the intensities for the three models.
The overall behaviour is consistent but the XY and Heisenberg models
manifests diffuse scattering at different Q ranges thus, by comparing these predictions with a real data set, one should be able to discriminate among these models.}
\end{figure*}

To confirm our understanding of the phonon density of states (DOS) and its relationship to the crystal field excitation spectrum, the phonon DOS of \LUCL{} was calculated using Density Functional Theory(DFT). The calculations were performed using the Vienna Ab initio Simulation Package (VASP)~\cite{VASP}. 
The calculation used Projector Augmented Wave (PAW) method~\cite{BlochlPE,Kresse} to describe the effects of core electrons, and Perdew-Burke-Ernzerhof (PBE)~\cite{Perdew} 
implementation of the Generalized Gradient Approximation (GGA) for the exchange-correlation functional. The energy cutoff was 600 eV for the 
plane-wave basis of the valence electrons. The lattice parameters and atomic coordinates determined by neutron diffraction at 10 K (see Tab.~\ref{tab: 0}) 
were used as the initial structure. The electronic structure was calculated on a $6\times3\times6$ $\Gamma$-centered mesh for the unit cell, 
and a $3\times3\times3$ $\Gamma$-centered mesh for the $2\times1\times2$ supercell. 
The total energy tolerance for electronic energy minimization was $10^{-8}$ eV, and for structure optimization it was 
$10^{-7}$ eV. The maximum inter-atomic force after relaxation was below $0.001$ eV/\AA, and the pressure after relaxation was below 2 MPa. The relaxed 
lattice parameters are $a=6.620$ \AA, $b=11.437$ \AA, $c=6.319$ \AA, $\alpha=90^\circ$, $\beta=110.395^\circ$, $\gamma=90^\circ$. A Hubbard U term of 6.0 eV 
was applied on Lu for the localized $4f$ electrons~\cite{Jiang}. The optB86b-vdW functional~\cite{Klimes} for dispersion corrections was used to describe the 
van der Waals interactions between layers. Force 
constants were calculated using Density Functional Perturbation Theory (DFPT) as implemented in VASP, and the vibrational eigen-frequencies 
and modes were then calculated by solving the dynamical matrix using $Phonopy$~\cite{Togo}. Non-analytical correction was applied to account for the LO-TO 
splitting~\cite{Gonze}. The $OClimax$ software~\cite{ChengYQ} was used to convert the DFT-calculated phonon results to the simulated INS spectra.

The calculated $S(\vec{Q},\omega)$, the relative DOS and the phonon dispersion across the highly symmetric directions
of the Brillouin zone for \LUCL{}, are shown in Fig.~\ref{fig: 4}. The DFT calculations are in very good agreement with the data  displayed in Fig.~\ref{ins_sup}(c) for
\LUCL; in particular we can see that the first and third excited states (represented with dashed lines) are located in a range of energies where phonons are totally absent effectively ruling out a possible hybridization. 

\section{Reverse Monte Carlo Analysis}

In order to extract information about the spin correlations in \YBCL{} from the powder data set we collected at HYSPEC, we performed a Reverse Monte Carlo (RMC) calculation following the approach described in Ref. [\onlinecite{Paddison2013}]. The program
$SpinVert$ was used to fit the temperature subtracted data set (as explained in the main text), integrated in the energy range $\hbar\omega$=[0.1,1.2] meV 
to avoid contamination from the elastic line. 

A simulation super-cell of $6\times6\times6$ containing a total of $6^4 = 1296$ spins with periodic boundary conditions (PBCs) has been
used to fit the spin excitation spectrum. For simplicity, the first attempt assumed all spins in the super-cell have Ising-like nature and are completely uncorrelated. The RMC has been done with a statistical average of 100000 configurations.  The only fitting parameter in the
code was a scaling factor to match the intensity of our calculation with the data set. The final spin configurations generated by the RMC were then averaged and the best fit with the HYSPEC data gave $R_{wp}=7.1$, and a $\chi^2=11.1$.

In order to verify our model we performed a campaign of RMC simulations changing the anisotropy of the spins to XY-like and/or Heisenberg, increasing the size of the simulation super cell
to study possible size effects, and repeating the same type of calculations in monoclinic coordinates to check differences in the diffuse scattering pattern. 
We found that increasing the size of the super-cell above $6\times6\times6$ does not affect the quality of the fit,
nor the calculated diffuse scattering; the calculation simply takes longer due to the higher number of spins in the super-cell. In the same way, performing the 
calculation in hexagonal or monoclinic notation does not affect the final result.   

The lack of a spin Hamiltonian that describes the exchange interactions prevents us from fully understanding the dynamics in \YBCL{} at this time. However, the analysis of the angular distribution of
the spins in the Heisenberg model (see Fig.~\ref{fig: 7}) indicates that the spins are preferentially laying in the $ab$ plane as suggested by the local anisotropy of the crystal field ground state. Finally, from the generated spin configurations, we can calculate and compare the radial spin-spin correlation function~\cite{Paddison2013} for all three models (i.e., the Ising-like, XY-like, and Heisenberg models), 
to analyze the spin-spin correlations. Figure~\ref{fig: 7}(b) shows that there is an antiferromagnetic nearest
neighbor spin correlation. The second and third neighbors are ferromagnetically correlated.  Furthermore, this behavior is qualitatively the same for all three models explored here.

The RMC calculations also provide a prediction of the diffuse scattering pattern which can be used to further constrain the type of spin-spin correlation when compared to single crystal neutron diffraction experiments.  The calculated diffuse scattering patterns are shown in Fig.~\ref{fig: 5}(b-c).


\end{document}